# Suppression of Octahedral Tilts and Associated Changes in Electronic Properties at Epitaxial Oxide Heterostructure Interfaces

A. Y. Borisevich,[*,1] H.J. Chang,[1] M. Huijben,[2,3] M.P. Oxley,[1] S. Okamoto,[1]

M.K. Niranjan,[4] J.D. Burton,[4] E. Y. Tsymbal,[4] Y.H. Chu,[5] P. Yu,[3]

R. Ramesh,[3] S.V. Kalinin,[1,] and S.J. Pennycook[1]

[1] Oak Ridge National Laboratory, Oak Ridge, Tennessee 37831

[2] Faculty of Science and Technology, MESA[+] Institute for Nanotechnology, University of Twente, P.O. BOX 217, 7500 AE, Enschede, The Netherlands

[3]Department of Materials Science and Engineering and Department of Physics, University of California, Berkeley, California, 94720

[4]Department of Physics and Astronomy, Nebraska Center for Materials and Nanoscience, University of Nebraska, Lincoln, Nebraska, 68588

[5]Department of Materials Science and Engineering, National Chiao Tung University, Hsinchu, Taiwan 30013 (ROC)

---

[*] albinab@ornl.gov

Epitaxial oxide interfaces with broken translational symmetry have emerged as a central paradigm behind the novel behaviors of oxide superlattices. Here, we use scanning transmission electron microscopy to demonstrate a direct, quantitative unit-cell-by-unit-cell mapping of lattice parameters and oxygen octahedral rotations across the $BiFeO_3$-$La_{0.7}Sr_{0.3}MnO_3$ interface to elucidate how the change of crystal symmetry is accommodated. Combined with low-loss electron energy loss spectroscopy imaging, we demonstrate a mesoscopic antiferrodistortive, phase transition and elucidate associated changes in electronic properties in a thin layer directly adjacent to the interface.

The unique electronic and dielectric properties of oxide heterostructures have recently propelled these systems to the forefront of condensed matter physics. Controlling charge transfer across the interfaces gives rise to spectacular behaviors including interface mediated conduction,[1] superconductivity,[2] magnetic effects[3] and improper ferroelectric responses.[4] However, the full spectrum of electronic behaviors at the ferroelectric-oxide interfaces that includes interface reconstructions and coupling to octahedral rotations, remains enigmatic and largely unexplored.

Here we utilize direct structural mapping[5,6] by high-resolution scanning transmission electron microscopy (STEM) to determine lattice parameters and oxygen octahedral rotations at an epitaxial $BiFeO_3$ - $La_{0.7}Sr_{0.3}MnO_3$ (BFO-LSMO) interface combined with electronic structure imaging by electron energy loss spectroscopy (EELS).[7] This allows us to directly correlate the atomic structure, polarization, strain fields and dielectric behavior on the atomic level. Combined with density functional calculations the data reveal the formation of a mesoscopic interface–induced phase at the BFO-LSMO interface such that a thin layer of BFO transforms into a phase with reduced band gap or possibly metallic behavior.

Multiferroic $BiFeO_3$ (BFO) was grown on a ferromagnetic $La_{0.7}Sr_{0.3}MnO_3$ electrode in the (100) orientation. Details of the epitaxial synthesis are reported in the supplementary section. While bulk rhombohedral $BiFeO_3$ is ferroelectric and antiferromagnetic, the presence of strong electric fields in major crystallographic directions[8] or pressure[9] could stabilize high-symmetry metallic (or low band gap) phases. This behavior has recently been explored by Bellaiche et al. by controlling the interplay of the antiferrodistortive order parameter and polarization. In BFO the octahedral tilts are related to the Fe-O-Fe angles and high/low spin states of Fe atoms, and hence directly control the metal-insulator transitions and magnetic properties.[10]

Here, we build an integrated picture of local structural distortions in the vicinity of the interface by directly measuring lattice parameter changes and octahedral rotation angles for every unit cell. To that end we use atomic position quantification from aberration corrected STEM images, building upon the approach recently introduced by Jia et al.[5,6,11] for TEM images. Shown in Fig. 1(a) is the high angle annular dark field (HAADF) STEM image of the 50 nm BFO/5nm LSMO/STO heterostructure (STO is outside of the field of view). The image shows clear contrast between the LSMO and BFO, suggesting that the interface is sharp and cationic mixing at the interface is minimal. The map in Fig.1 (b) is the 2D representation of the values of the pseudocubic lattice parameter $c$ (***normal*** to the interface) calculated from atomic column positions in (a); the increase from LSMO to BFO can be easily seen. The map shows a local increase in the pseudocubic $c$ parameter in the first 3-4 layers of BFO adjacent to the interface. Note that this increase is localized near the interface and hence cannot be accounted for by the usual volume-conserving Poisson's distortion due to epitaxial strain, which would act throughout the entire thickness of the film. The line plot in Fig. 1(c) shows that the lattice parameter in the first ~3-4 layers of BFO becomes as high as 4.21Å before settling down at a value of ~ 4.06Å after the first 4 layers, which is consistent with the value of 4.07 Å observed from X-ray diffraction data. There is no detectable change in the average value of the lattice parameter parallel to the interface, consistent with a high-quality, pseudomorphically constrained film. Identical behavior, up to and including the numerical value of the increased $c$ parameter, is observed in an ultrathin 3.2 nm BFO/5 nm LSMO/STO film (Fig. 1(d-f)).

To complement the analysis of the strains, we explore the octahedral tilts in the vicinity of the interface. We determine the tilt *angles* from the absolute positions of the Bi columns in two dimensions obtained from the HAADF STEM image in Fig. 1(d), and the oxygen columns obtained from a simultaneously-acquired bright field (BF) STEM image, given in Fig. 2(a). The two-dimensional tilt angle map in Fig. 2(b) shows a characteristic checkerboard pattern, consistent with the (001) projection of rhombohedral BFO (Fig. 2(c)). Extensive investigation of the potential errors due to tilt and defocus indicated that such systematic errors were negligible compared to the experimental statistical errors (Suppl. Mat.). The behavior of the tilt angles on even and odd sites in the lattice as a function of separation from the interface is shown in Fig. 2d The error bars are set to the standard deviation of the Bi-Bi angles. Notice that the absolute values of tilt angles increase gradually in the BFO phase over ~4-5 atomic layers, the same region that shows the expanded c-lattice parameter in Fig. 1(f). The saturated value of the tilt angle (~7°) far from the interface is approaching the value for bulk BFO (the projected $\theta_{BFO} \sim 10°$ in this viewing direction).

Insight into chemical composition and local dielectric properties of the interface comes from an EELS image of the heterostructure [Fig. 3(a)]. The energy range of -10 – 350 eV includes the zero-loss peak (around 0 eV) and core loss edges (Ti M, Mn M, Fe M, La N) at ≥35eV. In the intermediate energy range (~5 to 30 eV), the electron energy losses are dominated by plasmon generation, and hence contain information on the dielectric function, dielectric properties, and electron density.[7,12] A map of zero-loss intensity yields an energy filtered bright field image of the interface, allowing compensation for specimen drift.

Figure 3 (b) shows typical spectra (after Fourier-log deconvolution) from the three regions of the image and identifies different spectral features. From Fourier-log analysis, the sample thickness was quite uniform within the image (~50 nm), providing optimal conditions for low-loss imaging at which volume plasmonic excitations dominate. Clear differences are seen in both the plasmon energy and shape, and in the core-loss region, between the STO, LSMO, and BFO regions. To explore spatial variability of electronic structure, we utilize multivariate statistical analysis.[13] The 3D EELS data set in the selected energy interval is decorrelated using principal component analysis (PCA), representing the EELS spectra as a linear superposition of orthogonal eigenvectors with position dependent weight factors,

$$S_i(E_j) = a_{ik} w_k(E_j). \qquad (1)$$

where $w_i(E_j)$ are the eigenvectors and $a_{ik}(x,y)$ are position dependent expansion coefficients. The summation is performed over the first several statistically significant components chosen on an eigenvalue criterion.[14] The PCA loadings are then used as input for a linear neural network (NN) trained to recognize the individual components of the heterostructures. This approach is equivalent to the oblique PCA decomposition and hence does not introduce artifacts into the data. This analysis was also validated using multiple least-squares fitting (Suppl. Mat.)

The recognition analysis was carried out separately in three different energy ranges: -5 to 5 eV (zero-loss), 5 to 32eV (low-loss), and ≥35eV (core-loss); for the last two data sets the spectra were first treated via Fourier-log deconvolution of the zero-loss peak to remove multiple scattering and the zero-loss peak. The resulting recognition maps for different energy intervals are shown in Figs. 3(c-e). The zero-loss (energy-filtered bright field) image is dominated by mass-thickness contrast, which is not chemistry-specific, and the corresponding recognition image [Fig. 3(c)] does not separate the three individual components. Note that the ripples in the image correspond to the atomic periodicities in the underlying perovskite lattice (bright-field contrast).

The examination of the recognition map for the core-loss region [Fig. 3 (e)] shows clear contrast associated with individual phases, i.e. provides chemical identification of the components. The spatial resolution of this map as determined by the 25-75% criterion[15] is ~0.8 nm. It is independent of the number of PCA components (if 4 or more are used), indicating the fidelity of the data analysis, and is also consistent with the expected resolution of low-loss EELS. Note that due to close proximity of the Ti, Mn, and Fe M edges and plasmon peaks, the recognition problem is not accessible by standard background subtraction. Remarkably, in the low-loss energy range [Fig. 3(d)] corresponding to plasmon excitations, a region at the BFO/LSMO interface is clearly identified as a new phase that has electronic properties distinct from the constituent components. In other words, in this energy range the EELS spectra of the interface region cannot be interpreted as a linear superposition of constituents, suggesting the presence of a layer with anomalous dielectric properties (this conclusion is also supported by a multiple least squares fit, see Supplementary Materials). The width of the anomalous layer is ~2 nm (5 unit cells), above the resolution limit of low-loss EELS, and closely coincides with the width of the strained layer. This layer can be observed in the EELS images of both thick (50 nm) and thin (3.2 nm) films that have opposite polarization orientations, and can be identified using the cross-trained neural net. In comparison, the STO-LSMO interface is identified as abrupt (within the resolution) in all three energy ranges.

The comparison of EELS (anomalous layer) and structural (tilt angle, lattice expansion) data suggests that the origins of the observed behavior lie in the coupling of the antiferrodistortive order parameter (i.e. octahedral tilts) across the interface. Indeed, for ferroelectric materials the propagation of the zone boundary modes[16] and closure polarization patterns[17] across the interface has been demonstrated as an effective mechanism for stabilization of polarization in the ferroelectric-oxide heterostructures. Similarly, the coupling of the antiferrodistortive order parameter plays a crucial role in the STO-LaTiO$_3$ system[18,19] and the STO-LaAlO$_3$ system.[20] The continuity of the octahedral sublattice requires tilt angles to be continuous across the interface. Since the octahedral tilts in LSMO are much smaller (due to the proximity of STO substrate) and are close to zero at the interface (see Fig. S7 in Suppl. Mat.), the BFO tilt angle is suppressed at the interface. Note that relaxation of the tilt angle can proceed only through deformation of octahedra, i.e. direct change in the electronic structure of the central cation. Hence, the interfacial electronic property is expected to be greatly influenced by these structural changes. The increase of the Fe-O-Fe angle (towards 180 degrees) and forced higher symmetry is expected to increase the band-width of occupied and unoccupied bands, $W_O$ and $W_{UO}$, respectively, reducing the band gap in the BFO.

Density functional theory (DFT) calculations support this conjecture. We conducted both local spin-density approximation (LSDA) and LSDA+U calculations of the bulk BFO band structure.[21] In both cases, forcing BFO into tilt-free cubic symmetry significantly reduces the band gap. In the case of LSDA, the band gap changes from 0.40 eV to 0, and the ground state becomes ferromagnetic, rather than G-type antiferromagnetic (Fig 4(a)). For LSDA+U, we have chosen $U_{eff}$ = 3.0 eV, which gives a direct band-gap of 2.5 eV at the Γ-point, compared to the experimental direct band gap of 2.7 eV.[22] The calculated indirect- gap is reduced from 1.7 eV to 0.74 eV on suppressing the octahedral tilts; the ground state remains G-type antiferromagnetic and insulating [Fig.4(b)]. In both the LSDA and LSDA+U cases we find that the decrease in the band gap comes mainly from an enhancement of the band-width, especially of the unoccupied Fe minority-spin $d$-states. This is consistent with an increase in the Fe-O-Fe angle to 180° with suppressed octahedral tilting, which increases interatomic hopping and therefore the band-width.

Notably, the distortion of the BFO itself is not the only possible source of changes to the electronic structure at this interface. For instance, evanescent wave function tails from the LSMO penetrate into the BFO introducing a finite density of Metal Induced Gap States within the band gap.[23] The finite density of states (DOS) further enhances the conductive behavior and may be the reason that the octahedral tilts are pinned to near zero at the interface (Fig.4(c)).

These studies demonstrate the new paradigm of an interface phase transition mediated by the antiferrodistortive coupling across the interface, complementing the established polarization- and charge driven behavior. This also implies that novel ferromagnetic properties can arise in the vicinity of the interface as the result of the deformation of oxygen octahedra surrounding cation and changes in M-O-M angles.[24] These results illustrate that controlling the octahedral tilting behavior can provide a new, and virtually unexplored, dimension to the behavior of the ferroelectric and multiferroic films. Selectively tuning (either suppressing or enhancing) the tilts using the appropriate substrates can thus be used to establish new phases with novel properties, extending the concept of the strain- and polarization controlled interfaces to a broad new class of tilt-controlled phenomena.

ORNL and ORISE. The work at Berkeley is partially supported by the Semiconductor Research Corporation as well as by the US DOE under contract no. DE-AC02-05CH1123. The work at Nebraska is supported by the NSF-funded MRSEC (grant No. DMR-0820521), the Nanoelectronics Research Initiative of the Semiconductor Research Corporation and the Nebraska Research Initiative. Computations were performed utilizing the Research Computing Facility at UNL and the Center for Nanophase Materials Sciences at ORNL.

**Figure Captions**

**Figure 1.** Lattice parameters from HAADF STEM: (a) image of the 50nmBFO/5nmLSMO/STO thin film, (b) corresponding 2D map of the $c$ lattice parameter (perpendicular to the interface), with lighter color corresponding to higher $c$ values, (c) line profile obtained by averaging all rows of the map in (b) showing an increase in the first few layers of BFO. (d) Image of the 3.2nmBFO/5nmLSMO/STO ultrathin film, (e) corresponding 2D map of the $c$ lattice parameter, (f) line profile obtained by averaging all rows of the map in (e). Image in (d) was rescaled to correct for drift; error bars in (c) and (f) reflect variance between individual rows in the 2D maps. Note that graphs (c) and (f) are presented on the same absolute scale. Scale bars are 1 nm.

**Figure 2.** Oxygen octahedral rotation angles from BF STEM: (a) image of the 3.2nmBFO/5nmLSMO/STO ultrathin film, acquired simultaneously with the HAADF image in Fig 2(d), (b) corresponding 2D map of in-plane octahedral rotation angles in BFO showing checkerboard order, (c) BFO structure in rhombohedral (001) orientation showing the tilt pattern, (d) line profile obtained from the map in (b); two bad points in the map corresponding to a hole in the sample are taken out of the averaging. The image in (a) was rescaled to correct for drift; the map in (b) was corrected for local Bi-Bi angle variations; error bars in (d) are equal to the standard deviation of local Bi-Bi angles. Scale bar is 1 nm

**Figure 3.** Low loss EELS imaging of the $SrTiO_3/(La,Sr)MnO_3/BiFeO_3$ interface. (a) imaging area; (b) representative spectra (after Fourier-log deconvolution) of the three components

showing distinctive signatures and features; (c-e) resulting recognition maps for (c) zero loss region (-5 to 5 eV), (d) plasmon excitation region (5 to 30 eV), and (e) low-lying core edges (35 to 150 eV). Note that a region of BFO approximately 2 nm wide is not recognized as any of the components in map (d). Scale bar is 2 nm.

**Figure 4.** Interfacial lattice distortion in BiFeO$_3$ and its effect on the density of states. (a) DOS by LSDA method, in fully relaxed BiFeO$_3$ (top) where oxygen octahedron rotations are present and (bottom) in cubic BiFeO$_3$ where oxygen octahedron rotations are absent; band gap is reduced from 0.44 eV to 0 eV. (b) DOS by LSDA+U method, U-J= 3.0 eV (top) in fully relaxed BiFeO$_3$ where oxygen octahedron rotations are present and (bottom) in cubic BiFeO$_3$ where oxygen octahedron rotations are absent; band gap is reduced from 1.7 eV to 0.74 eV. The dashed lines indicate the position of the Fermi energy. (c) Additional coupling with LSMO can generate finite density of states within the gap.

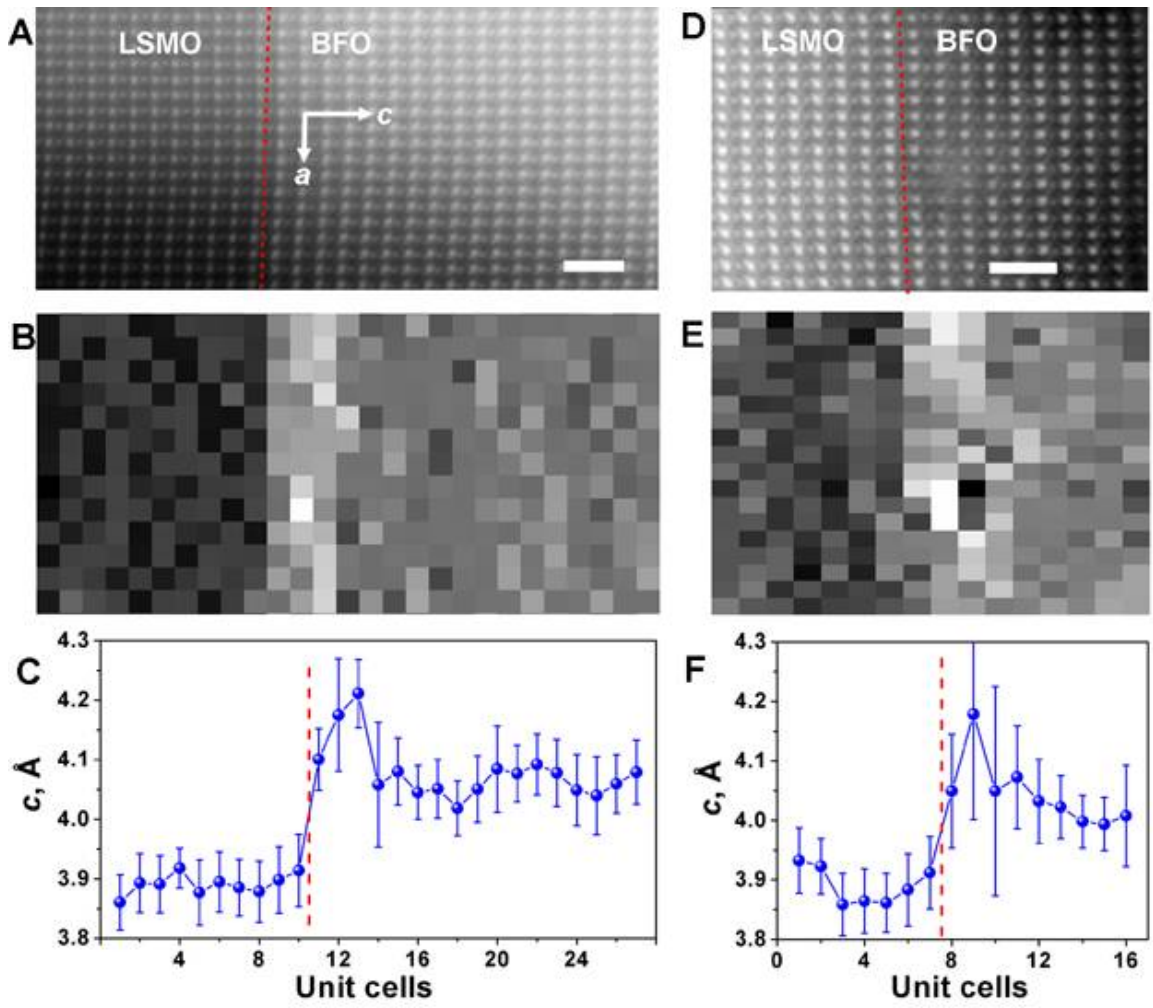<000a><000a><000a><000a><000a><000a><000a><000a><000a><000a><000a><000a><000a><000a><000a><000a><000a><000a><000a><000a><000a><000a><000a><000a><000a><000a><000a><000a><000a><000a><000a><000a><000a><000a><000a><000a><000a><000a><000a><000a><000a><000a><000a><000a><000a><000a><000a><000a><000a><000a><000a><000a><000a><000a><000a><000a><000a><000a><000a><000a><000a><000a><000a>

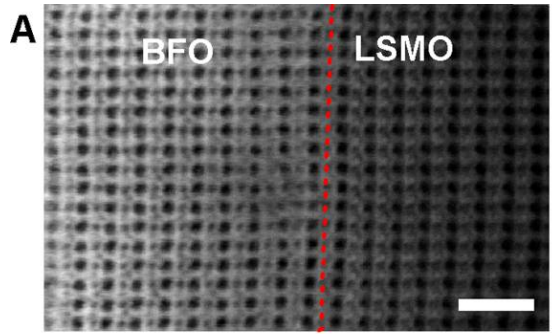

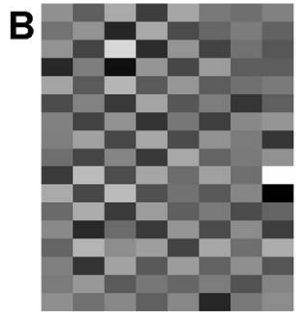
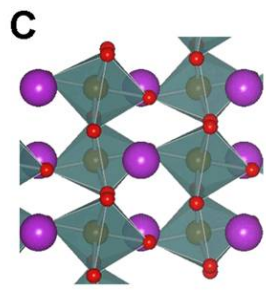

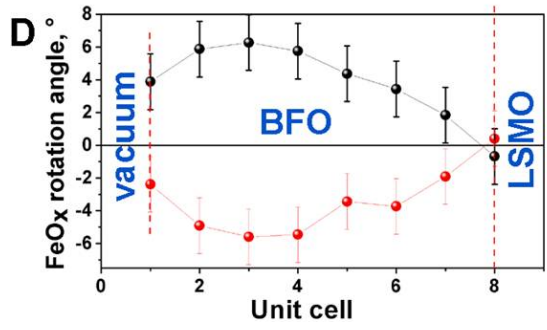

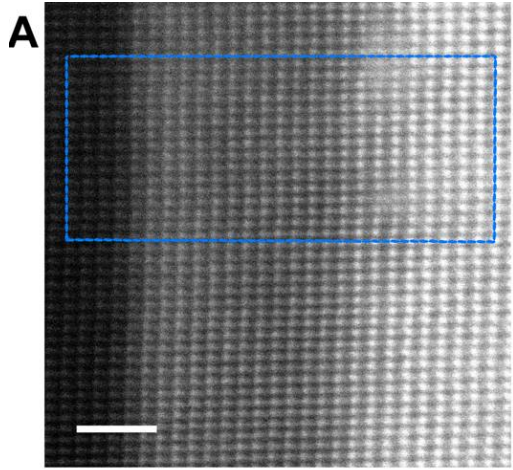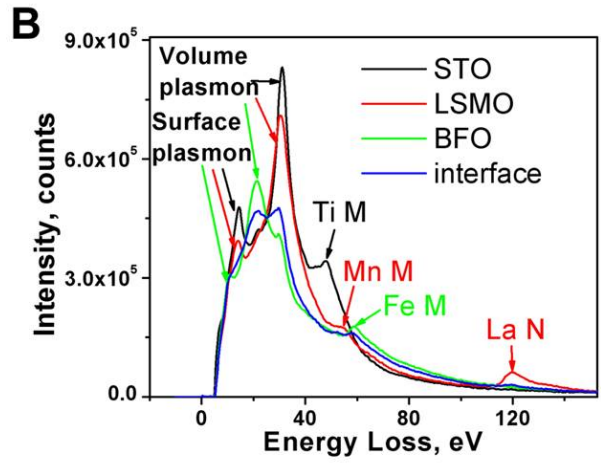
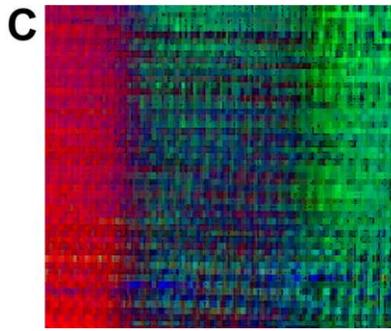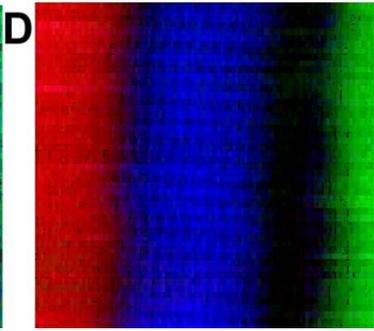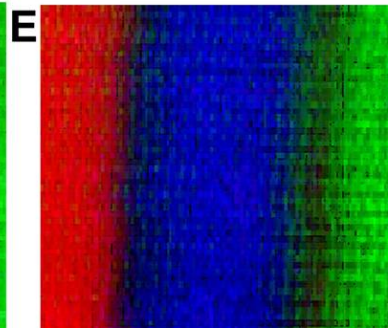

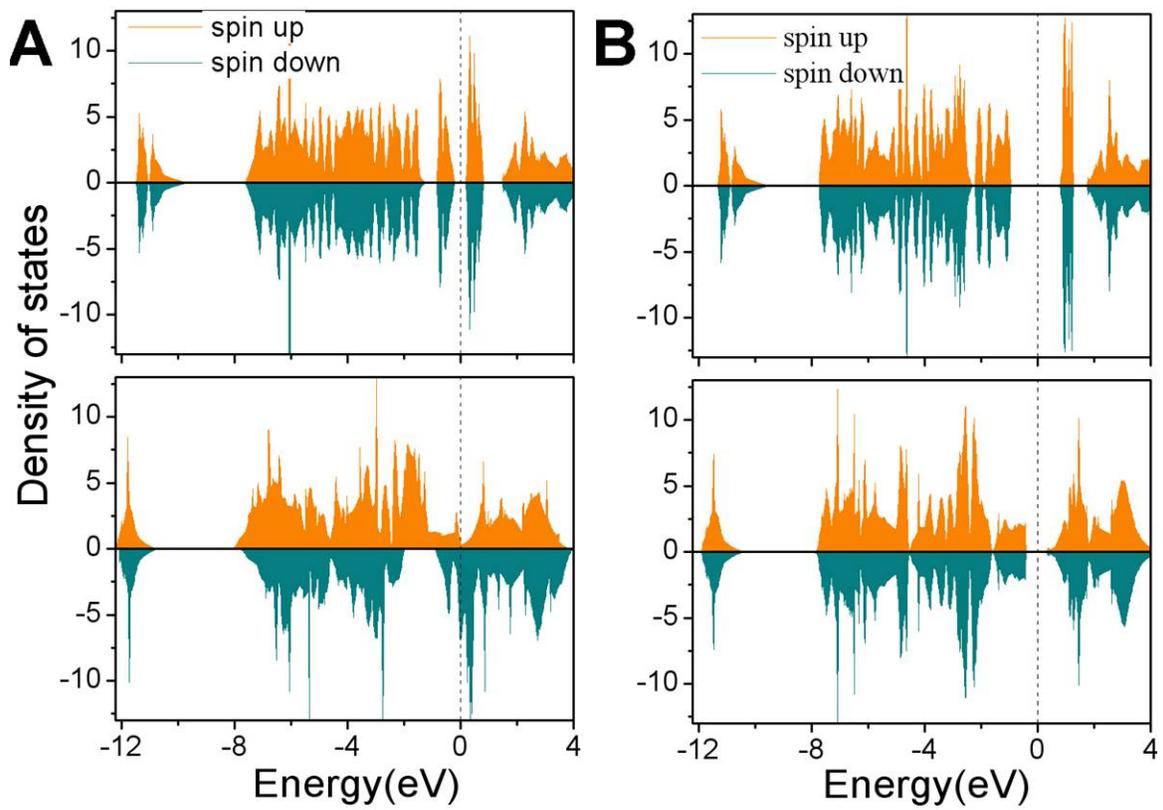
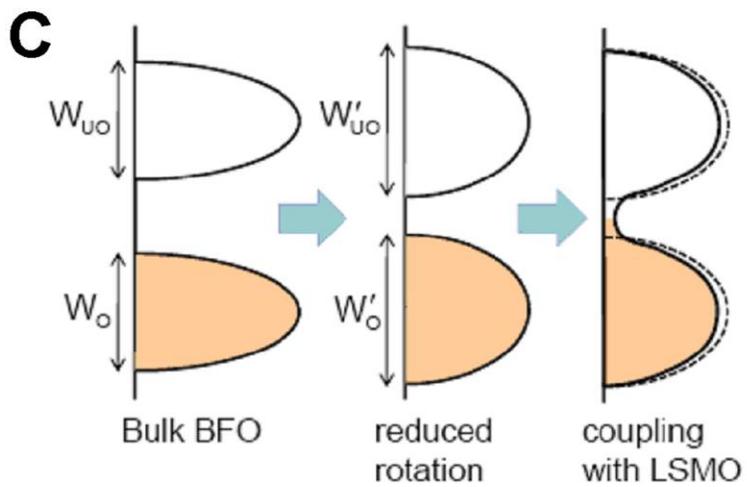

*Supplementary Materials for the manuscript*

**Suppression of Octahedral Tilts and Associated Changes of Electronic Properties at Epitaxial Oxide Heterostructure Interfaces**

A. Borisevich,*,[1] H.J. Chang,[1] M. Huijben,[2,3] M.P. Oxley,[1] S. Okamoto,[1]

M.K. Niranjan,[4] J.D. Burton,[4] E.Y. Tsymbal,[4] Y.H. Chu,[5] P. Yu,[3]

R. Ramesh,[3] S.V. Kalinin,[1] and S.J. Pennycook[1]

[1] Oak Ridge National Laboratory, Oak Ridge, Tennessee 37831

[2] Faculty of Science and Technology, MESA+ Institute for Nanotechnology, University of Twente, P.O. BOX 217, 7500 AE, Enschede, The Netherlands

[3] Department of Materials Science and Engineering and Department of Physics, University of California, Berkeley, California, 94720

[4] Department of Physics and Astronomy University of Nebraska, Lincoln, Nebraska, 68588

[5] Department of Materials Science and Engineering, National Chiao Tung University, Hsinchu, Taiwan 30013 (ROC)

## 1. Synthesis and XRD details

BiFeO$_3$ (BFO) - La$_{0.7}$Sr$_{0.3}$MnO$_3$ (LSMO) heterostructures were fabricated by pulsed-laser deposition with reflection high-energy electron diffraction (RHEED) control of the growth process. Atomically smooth TiO$_2$-terminated SrTiO$_3$ (STO) (100) substrates were prepared by a combined HF-etching/anneal treatment. All substrates had vicinal angles of ~0.1°. Stoichiometric LSMO and BFO targets were ablated at a laser fluence of ~1.5 J cm$^{-2}$ and a repetition rate of 1 or 2 Hz for the growth of LSMO and BFO, respectively. During growth, the substrate was held at 750°C, in an oxygen environment at 200 mTorr for LSMO[1], while for BFO the conditions were adjusted to 670°C and 100 mTorr [2]. RHEED analysis demonstrated intensity oscillations indicating layer-by-layer growth without any island formation. After the growth, the heterostructures were slowly cooled to room temperature in 1 atm of oxygen at a rate of ~5°C/min to optimize the oxidation level.

The crystal properties of the BFO-LSMO heterostructures were analyzed by X-ray diffraction to determine the epitaxial relation between the STO (001) substrate and the LSMO & BFO layers on top. Various BFO-LSMO heterostructures were studied with different BFO layer thicknesses in the range 0 - 50 nm, while the LSMO layer was kept at a constant thickness of 5 nm. The LSMO layer as well as the BFO layer was in-plane strained to the STO substrate, resulting in an in-plane lattice parameter of ~3.90 Å. This led to a tensile strain of ~0.8 % in the LSMO layer (bulk parameter ~3.87 Å) and subsequently a compressive strain of ~2.0 % in the BFO layer (bulk parameter ~3.98 Å). For the 50 nm thick BFO layer clear Kiessig fringes were observed, indicating a highly ordered crystalline BFO layer with a very smooth interface to the LSMO layer and a smooth top surface. From the peak position the out-of-plane BFO lattice parameter was determined to be ~4.07 Å, which is much larger than the bulk value of ~3.98 Å, due to the in-plane compressive strain in the BFO layer. The same out-of-plane lattice parameter was also measured for heterostructures with a 10 nm and 25 nm thick BFO layer. For thinner BFO layers of 5 nm and lower a double peak was observed, indicating the presence of two layers with different out-of-plane lattice parameters. A rough estimate of the out-of-plane lattice parameters of these two layers for the very thin 3.2 nm BFO layer is ~4.14 Å and ~4.00 Å.

## 2. STEM-EELS

The EELS data was acquired using a VG Microscopes HB603U operated at 300 kV equipped with a Nion aberration corrector and Gatan Enfina® spectrometer. The EELS image was acquired on a 51x120 spatial pixel grid with ~1Å/pixel. In strain and tilt analysis, images in Fig. 2(d) and 3(a) were first drift-corrected to force the average Bi sublattice into ideal rectangular arrangement.

Note that specimen drift during the longer exposure in STEM is not a limiting factor in these studies, since uniform drift can be effectively compensated for. In practice we find that the variance introduced by non-uniformity of the drift is much less than the intrinsic changes related to a heterointerface or defect. At the same time, a major advantage of STEM compared to TEM comes from being able to use thicker samples for lattice parameter mapping, thus avoiding depolarization field effects on ferroelectric phase stability that can be very significant for thicknesses of less than 5 nm, often reported for HRTEM samples. Mapping of oxygen positions and thus octahedral tilts in this paper is conducted using bright field (BF) STEM and is thus subject to the same sample thickness constraints as TEM determination.

### 2.a. Calibration of the lattice parameters

In STEM, global calibration can not be employed due to varying drift conditions and (in the case of the VG Microscopes HB603U) small magnification differences between images. While theoretically the best case scenario, calibration within each image was also not possible. To accurately determine atomic positions, we need ~30-40 pixels per unit cell, and that restricts the field of view to either primarily LSMO and BFO, or primarily LSMO and STO. Thus, a two-step approach was employed. The pseudocubic lattice parameter $c$ normal to the interface was calibrated by assuming that for STO it was equal to the bulk value, 3.905Å. From that value, the lattice parameter for LSMO was evaluated. For the several images examined, the confidence intervals for the LSMO lattice parameter overlapped; the value that was used for further calibration, 3.891Å, corresponded to the smallest confidence interval. The corresponding calibration graph is given in Fig. S1. The calculated value for LSMO was then used for calibration of the lattice parameters in each image of LSMO and BFO.

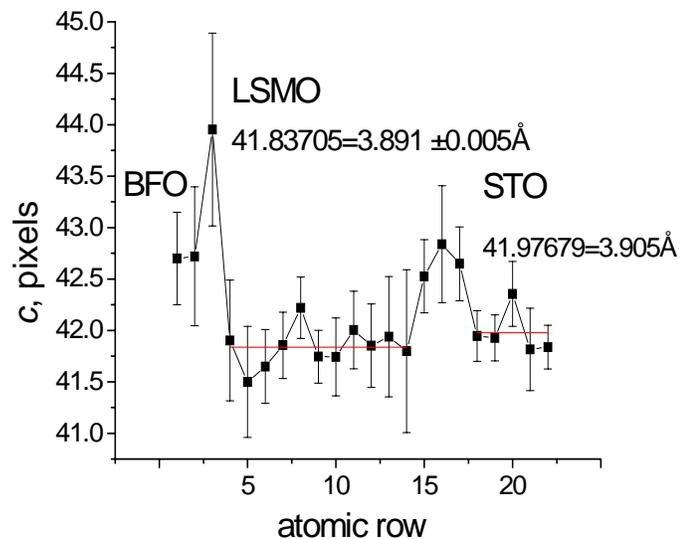

**Fig. S1**. The calibration graph used to determine lattice parameter normal to the interface.

**2.b. Recognition fit of EELS data**

The analysis of 3D spectral imaging EELS data can be based either on deconvolution methods, or multivariate statistical analysis. The Kramers-Kronig analysis on the spectrum image does not provide a clear picture of the interfacial phenomena since the real part of the calculated dielectric function does not reveal statistically significant variations of the crossover energy (the high-frequency dielectric constant), while the imaginary part shows local variations in the fine structure largely similar to those in the original spectrum. However, in this system and many other complex oxide systems, small changes in electronic structure, which can be more subtle than changing the overall value of dielectric constant, can nevertheless strongly affect the resulting properties. Here, we utilize *linear* multivariate statistical analysis to explore interface behavior. However, note that the developed framework can also be extended to non-linear recognition analysis.

**Principal component analysis.** The spectroscopic image of NxM pixels formed by spectra containing P points is represented as a superposition of the eigenvectors $w_j$, $S_i(E_j) = a_{ik} w_k(E_j)$, where $a_{ik} \equiv a_k(x,y)$ are position-dependent expansion coefficients, $S_i(E_j) \equiv S(x,y,E_j)$ is the image at a selected time, and $t_j$ are the discrete times at which response is measured. The eigenvectors $w_k(E)$ and the corresponding eigenvalues $\lambda_k$ are found from the covariance matrix, $\mathbf{C} = \mathbf{A}\mathbf{A}^T$, where $\mathbf{A}$ is the matrix of all experimental data points $\mathbf{A}_{ij} = S_i(E_j)$, i.e. the rows of $\mathbf{A}$ correspond to individual grid points ($i = 1,..,N \cdot M$), and columns correspond to time points, $j = 1,..,P$. The eigenvectors $w_k(t_j)$ are orthogonal and are chosen such that corresponding eigenvalues are placed in descending order, $\lambda_1 > \lambda_2 > ....$ . The eigenvalues and eigenvectors are determined through singular value decomposition of the **A** matrix (using the svd function of MatLab).

The anomalies in the BFO layer near the interface can be noticed in the PCA component maps above 4$^{th}$ order (which we use as input for neural network processing).

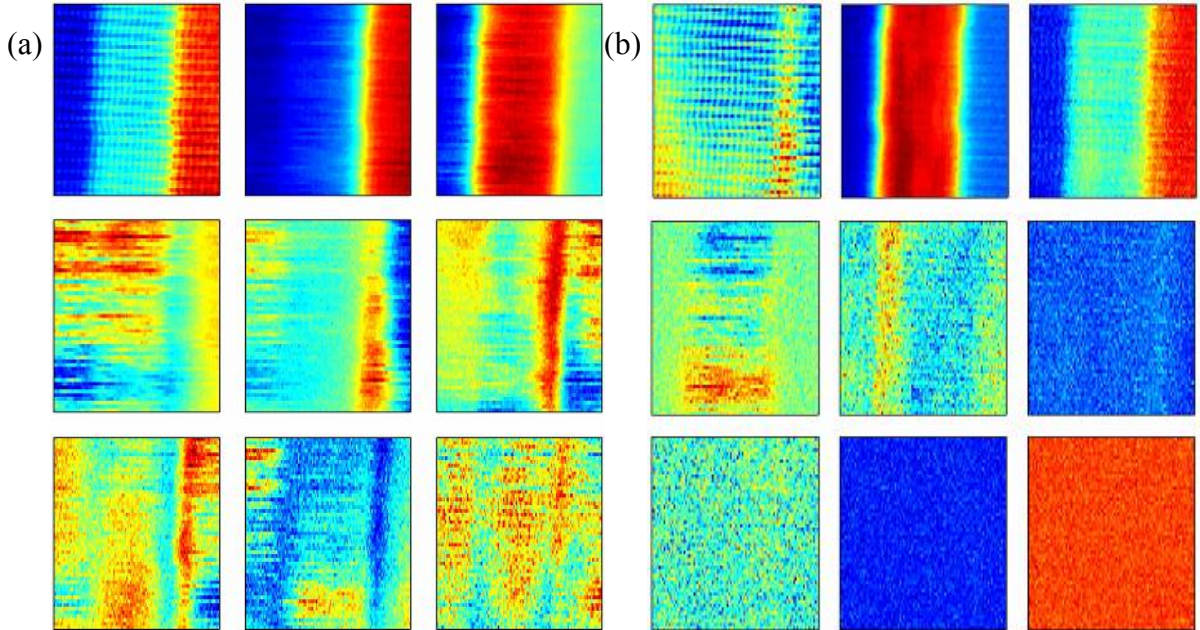

**Fig. S2.** First 9 PCA component eigenvalue maps for (a) low-loss range of 5-35 eV and (b) core-loss range of 35-125 eV.

Eigenvalue maps (1-9) for low-loss region are given in Fig. S2 (a). Note the anomaly visible in components 4-8. For comparison, eigenvalue maps (1-9) for the core-loss region are given in Fig. S2 (b). No anomaly is visible inside the BFO.

**Neural net fit.** To perform the mapping, subsets of the spectra far from the interface were identified with specific phases and used as a training set for a linear feed-forward neural network (5, 10, 3 neurons, linear transfer function) trained using back propagation. The trained network was then used to process the full experimental image. The software is implemented using the Neural Network Toolbox for MatLab.

## 2.c. Multiple least squares fit of the spectrum image.

Similar results to the ones reported in the paper using neural network analysis with respect to spatial localization of the discrepancies from a simple superposition model can be obtained from a multiple linear least squares (MLS) fit. Starting from the Fourier-log deconvolved spectrum, we take one spectrum for each component (averaged over the first column, the 120$^{th}$ column, and the 56$^{th}$ column out of 120 to represent STO, BFO, and LSMO respectively) (Fig. S3).

We then perform MLLS fitting in two energy ranges: 4.1 eV to 35 eV and 35 eV to 125 eV. Figure S4 gives the RGB-colored composites of the fit coefficients (green for STO, blue for LSMO, and red for BFO), residuals, and $\chi^2$ maps. Clearly, the region in BFO adjacent to the interface is not described well by a linear superposition model in the 4.1 to 35 eV energy range, supporting the analysis of the recognition fit.

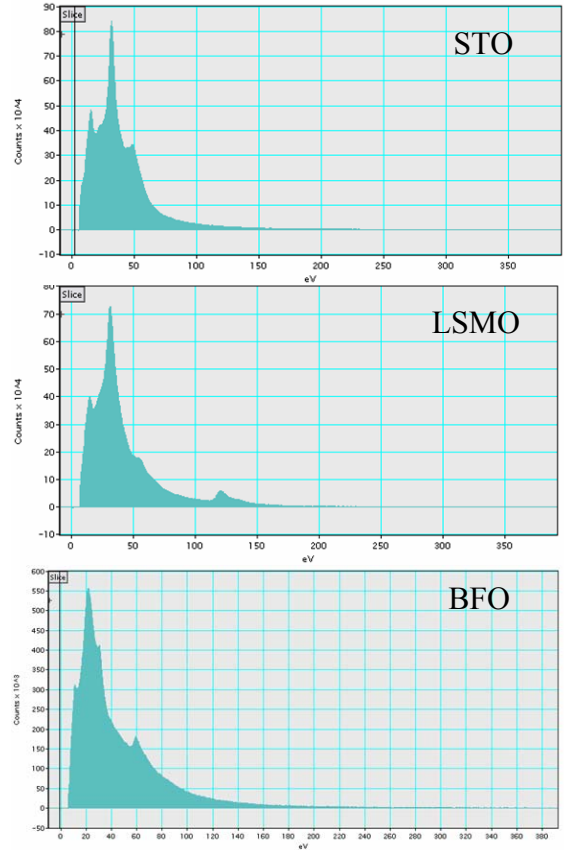

**Fig. S3.** Spectra (after Fourier-log deconvolution) of the individual components used in the MLS fitting analysis approach used in the paper.

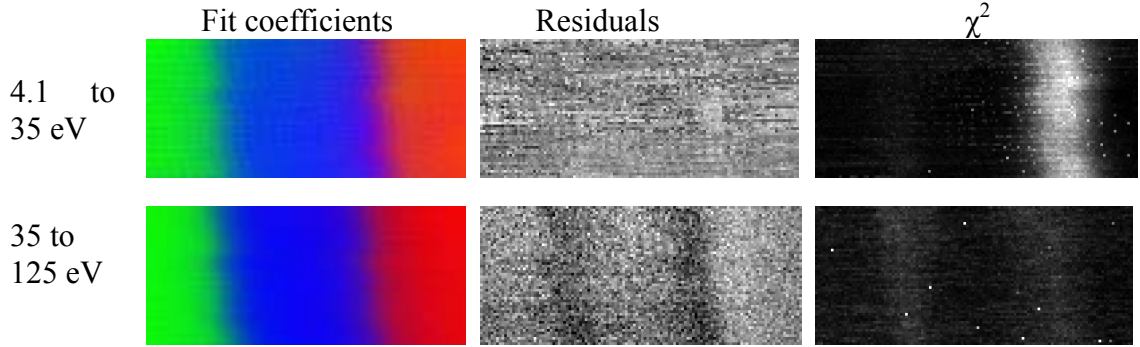

**Fig. S4.** Results of the MLLS fit performed with spectra in Fig. S3: RGB-colored composites of the fit coefficients (green for STO, blue for LSMO, and red for BFO), residuals, and $\chi^2$ maps for the two energy ranges.

While these alternative methods offer the same proof of a dielectric anomaly as the chosen method of PCA+NN recognition, and may be simpler in realization, we believe that the neural network approach, while equally valid statistically, gives advantages in both implementation and interpretation. For example, with some prior information about the system we could choose a non-linear recognition model. Even within the confines of the linear model, we can determine which PCA components, and in what ratios, contribute to the anomalous region, and can thus generate the "anomaly" spectrum which we can then attempt to interpret with the aid of theoretical modeling. The possibility of using a cross-trained neural network also enables us to recognize anomalies in systems where it is hard to find a reference for bulk-like behavior, such as ultrathin films and multilayers.

**2.d. HAADF and BF STEM image simulations for BFO.**

The calculations are based on the Bloch wave method outlined in Ref. [3]. All ADF calculations assumed an ADF detector spanning 60—300 milliradians. The BF detector was assumed to be 2 milliradians semiangle. A probe forming aperture of 23 milliradians was used and C5 is set at 100mm. The specimen thickness in all cases was assumed to be 5 nm. The focal series of bright field and dark field images were calculated starting at a $\Delta f = -190$ Å (slice 0) with 10 Å increments. The final defocus value (slice 39) is 200 Å. All images are 15 Å square, with a pixel size of 0.1 Å. Due to an inherent uncertainty in residual 3rd order aberration measurement, simulations were performed for several values of $C_s$.

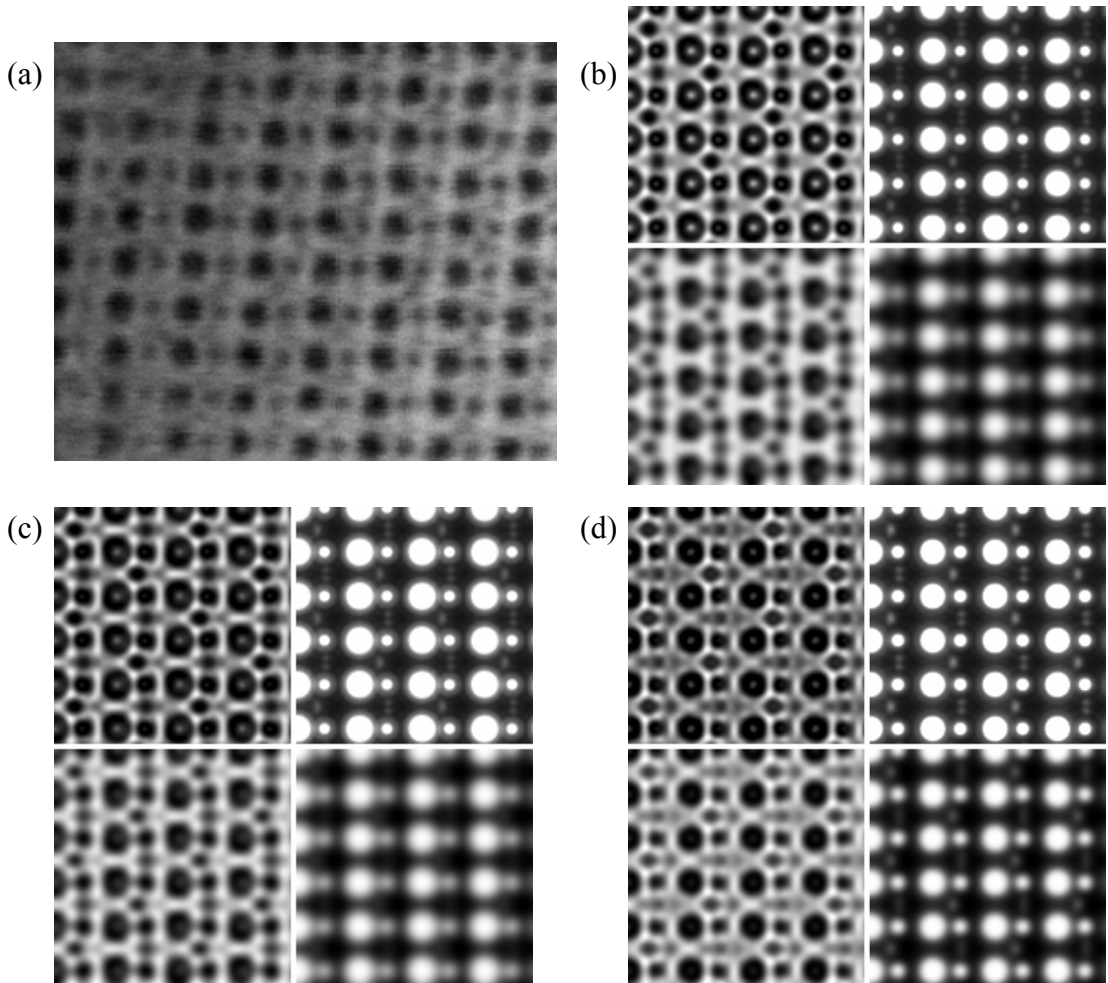

**Fig. S5.** (a) Raw experimental BF image. (b-d) Simulated BF (top right) and ADF (top left) images for $C_s$=-0.03 mm and $\Delta f$ = -50 Å (b), -0.037 mm and $\Delta f$ = -40 Å (c) and -0.05 mm and $\Delta f$ = 0 Å (d). Bottom rows of the panels show simulated images after Gaussian blur to account for spatial incoherence.

In all cases, in the conditions approximating experiment (i.e. all atoms black in bright field and a reasonably well resolved dark field image) oxygen (and other) atomic column images were centered exactly at the atomic positions. Small variations of contrast and thickness around these conditions did not move the atomic columns in either dark or bright field, although some contrast reversals occurred in BF. On the basis of the simulations, we conclude that determination of actual tilt angles from the tilt angles measured from BF STEM images is valid.

## 2.e. Control experiments for octahedral tilts

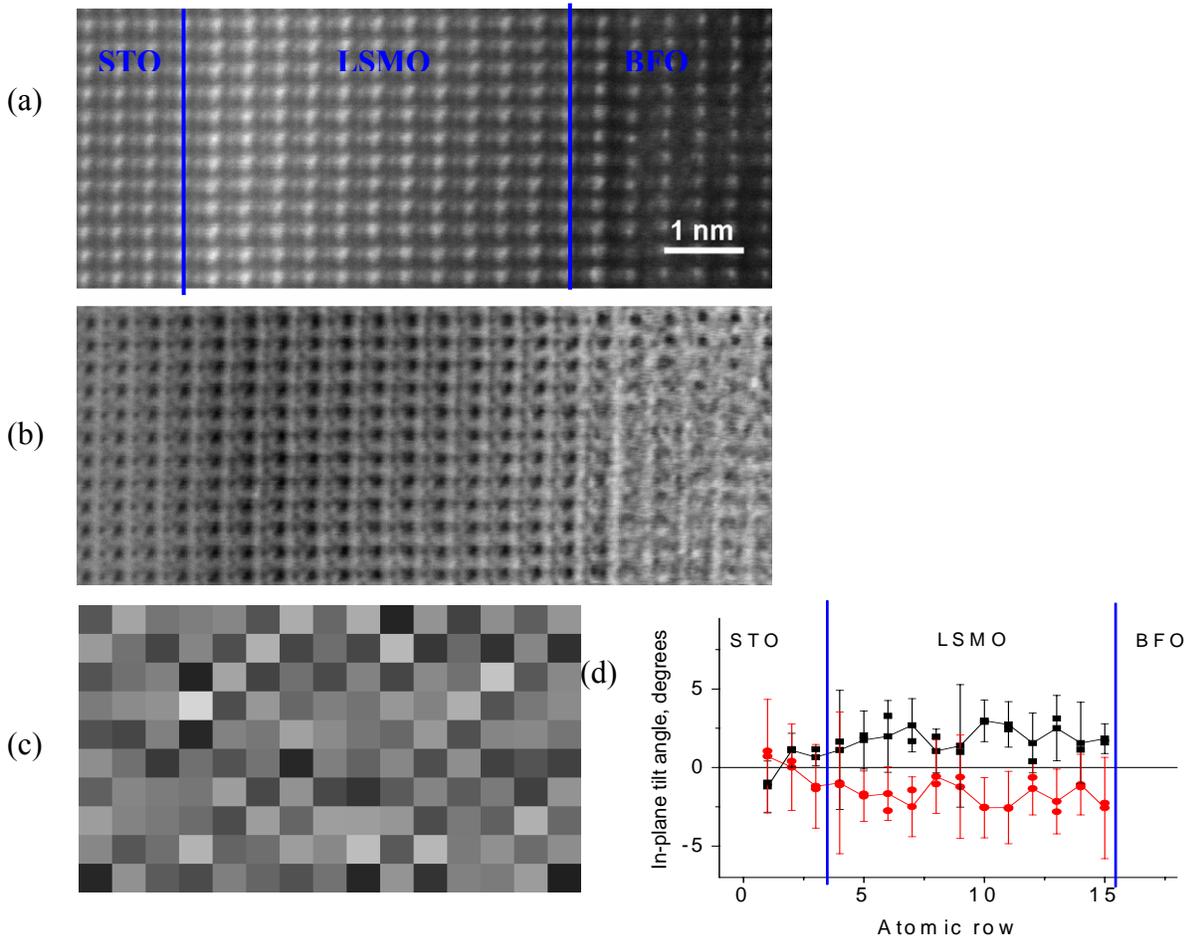

**Fig. S6.** Simultaneously acquired ADF(a) and BF(b) images of the STO-LSMO-BFO interface. (c) Tilt angle map generated from atomic positions in (a), (b). (d) Line profile of the map in (c) showing even and odd sites of the checkerboard lattice. There are no measurable tilts in STO, but the LSMO shows small tilts, consistent with its bulk rhombohedral structure.

The octahedral tilts were calculated from the tilt angles of the lines connecting adjacent oxygen columns with respect to the vertical. To compensate for local image distortions, the tilt angles of the lines connecting the nearest Bi atoms with respect to the horizontal were subtracted from the result. The correction resulted in better centering of the positive and negative branches around zero tilt

It was not possible to find a single image where octahedral tilts could be determined in STO, LSMO, and BFO simultaneously, because observation of oxygen in BFO and other materials required different focus conditions. Nevertheless, several images showed oxygen columns across the STO-LSMO interface and were quantified in a manner similar to that presented in the paper for BFO Fig. S6). It is seen that we can still measure the rhombohedral distortion in LSMO, where the tilt angle approaches 3 degrees in the center of the film. In STO, there are no apparent tilts, as expected.

**3. Theoretical methods.**

Density functional calculations were performed using the projector augmented wave method as implemented in the Vienna Ab Initio Simulation Package (VASP).[4] The Local Spin-Density Approximation (LSDA) and LSDA+U were used for exchange-correlation potential along with a plane wave basis set with a kinetic energy cutoff of 520 eV. A 6×6×6 Monkhorst-Pack mesh was used for k-point sampling and structures were relaxed until the ionic forces reduced to less than 0.005 eV/Å. The calculations did not include the spin-orbit interaction. Within the LSDA+U[5] approach the strong Coulomb repulsion between localized d states was treated by adding a Hubbard term to the effective potential, leading to an improved description of correlation effects. The LSDA+U method requires two parameters, the Hubbard parameter U and the exchange interaction J. Here we implement a simplified approach assuming a rotational-invariant potential which is described by an effective Hubbard parameter $U_{eff} = U - J$. Our results are consistent with the previous studies on the electronic structure of bulk BFO.